 \documentclass{vgtc}                          

\graphicspath{{figures/}{pictures/}{images/}{./}} 

\usepackage{times}                     

\usepackage{tabu}                      
\usepackage{booktabs}                  
\usepackage{lipsum}                    
\usepackage{mwe}                       

\usepackage{mathptmx}                  

\onlineid{5760}

\vgtccategory{Research}

\vgtcinsertpkg

\usepackage{comment}                   
\usepackage{stfloats}

\title{Ping! Your Food is Ready: Comparing Different Notification Techniques in 3D AR Cooking Environment}

\author{Aditya Raikwar\thanks{e-mail: adirar@colostate.com}\\ %
     \scriptsize Colorado State University %
\and Lucas Plabst\thanks{e-mail: lucas.plabst@colostate.edu}\\ %
     \scriptsize Colorado State University %
\and Anil Ufuk Batmaz\thanks{e-mail: ufuk.batmaz@concordia.ca}\\ %
     \scriptsize Concordia University \\
\and Florian Niebling\thanks{e-mail: florian.niebling@th-koeln.de}\\ %
     \parbox{1.8in}{\scriptsize \centering Advanced Media Institute, TH Köln \\ University of Applied Sciences, Köln, Germany}
\and Francisco R. Ortega\thanks{e-mail: fortega@colostate.edu}\\ %
     \scriptsize Colorado State University }

\teaser{
  \centering
  \includegraphics[width=0.9\linewidth, alt={Left: Notifications on Dock - All visual notifications are displayed as a list at the same position; Right: Notification on Object - Visual notifications are displayed above the objects that prompted them}]{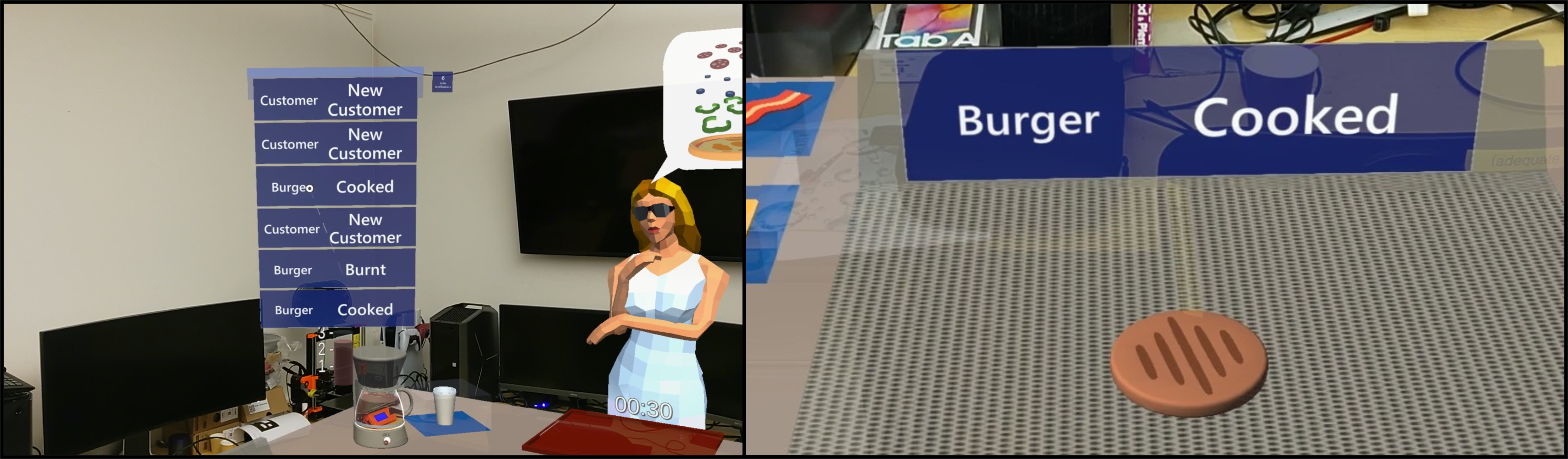}
  \caption{Left: Notifications on Dock; Right: Notification on Object}
  \label{fig:teaserNotification}
}

\abstract{
    Implementing visual and audio notifications on augmented reality devices is a crucial element of intuitive and easy-to-use interfaces. In this paper, we explored creating intuitive interfaces through visual and audio notifications. The study evaluated user performance and preference across three conditions: visual notifications in fixed positions, visual notifications above objects, and no visual notifications with monaural sounds. The users were tasked with cooking and serving customers in an open-source Augmented-Reality sandbox environment called ARtisan Bistro. The results indicated that visual notifications above objects combined with localized audio feedback were the most effective and preferred method by participants. The findings highlight the importance of strategic placement of visual and audio notifications in AR, providing insights for engineers and developers to design intuitive 3D user interfaces.
} 


\begin{document}


\firstsection{Introduction}

\maketitle

Interruptions are a significant research topic in Human-Computer Interaction (HCI), as humans have an innate ability to multitask \cite{koch2018cognitive}, leading to frequent switching between tasks by choice or due to external interruptions. This multitasking behavior is common in day-to-day life, such as checking emails during meetings. Gould et al. discuss that interruptions can have both positive and negative effects on user attention and task performance~\cite{gould2012multitasking}. While interruptive notifications can disrupt attention and hinder performance, well-designed notifications can aid in task management and reduce user anxiety by providing timely reminders and updates, particularly in dynamic and time-sensitive situations~\cite{iqbal2010notifications}.

Notification windows are commonly used to interrupt users in Head Mounted Displays (HMDs) applications \cite{rzayev2020effects}. When designing these notifications, it is essential to avoid obstructing the user's field of view (FoV), particularly during locomotion. In Augmented Reality (AR) HMD systems, designers must consider factors such as the placement of information, acceptable levels of occlusion, timing of the display of information appropriately, and selecting the most effective interaction method for users~\cite{imamov2020display}. Since notifications in AR-HMDs often involve displaying information directly within the user's visual field, conventional frameworks used for smart and desktop devices may not be directly applicable to the presentation of information in AR-HMD systems.

As display systems continue to advance, they are becoming increasingly integrated into users' daily lives. Screens have evolved from desktops to laptops, then to handheld devices, and now AR HMDs are placing screens optimized for clear and sharp images directly in front of our eyes. Moreover, the integration of virtual elements into the user's real-world environment through AR HMDs offers a seamless and immersive experience. All these advancements suggest that AR ``glasses'' are likely to become ubiquitous in the near future. Therefore, understanding the potential integration of these AR devices into daily life is crucial.

Understanding the user's context is crucial for delivering relevant information and controlling the pervasive devices~\cite{grubert2016towards}. Exploring the placement of notifications in an AR-HMD during demanding tasks, such as cooking and serving food to customers, is, to our knowledge, novel, particularly when considering the incorporation of both visual and audio modalities for notifications. The objective of this paper is to investigate how users can smoothly transition between tasks, guided by the Multiple Resource Theory (MRT)~\cite{wickens2022applied}, \textit{using visual and audio modalities}. The MRT offers insight into how users manage concurrent tasks and interruptions through different channels.

In this study, we looked at how notifications can help with the primary task as opposed to interrupting them. Previous studies have looked at the negative effects of interruptions on the primary task due to notifications. Studies like~\cite{upshaw2022hidden, ohly2023effects, kim2016analysis,  kushlev2016silence} state that task-independent notifications will decrease the performance. In this paper, we designed two types of world-fixed visual notifications: Notifications On Dock and Notifications On Object, as shown in~\autoref{fig:teaserNotification}. The audio notification may or may not accompany the visual notifications in the form of bubble popping sound (refer to the supplementary video for an example). To enhance participants' visibility and attention, we incorporated sound feedback. Our system was developed using the open-source ARtisan Bistro environment~\cite{raikwar2022artisan}. Various parameters in ARtisan Bistro, such as customer frequency and cooking speed, were adjusted to induce greater task-related stress, particularly regarding the participants' required task completion speed. Notifications were added to assist in task completion when user attention was interrupted. Our results showed that placing notifications directly on relevant objects can improve task performance and is preferred by users. Additionally, we demonstrated that the use of sound is an important factor in capturing the attention of users. Audio notifications enhanced the noticeability of visual cues, leading to better performance and higher user satisfaction. These findings highlight the importance of combining visual and audio notifications to optimize user experience in AR applications.

\section{Related Work}
\label{sec:relatedWork}

Unlike conventional smart devices, AR HMDs represent a novel computing environment, which poses unique challenges for implementing notification mechanisms. Understanding how to effectively utilize this new approach to deliver AR notifications on HMDs in real-world scenarios is crucial~\cite{paul2015interruptive, lee2022investigating}.

\subsection{Notifications}

One of the ways of getting users' attention is by means of notifications. Notifications can be delivered through visual, audio, and haptic mediums, and they generally fall into two categories~\cite{salazar2015notification}. The first category is \textbf{action-required notifications}, where the user is required to take some immediate action based on the information provided by the notification. E.g., Windows asking for administrative authentication. The second category is \textbf{passive notifications}, where the user is presented with some information that does not require the user to act, e.g., calendar notification.

A typical multimodal notification, such as receiving an email on a smartphone, includes a visual pop-up, a sound cue, and a vibration sequence to alert the user in various situations. Notifications vary based on context and user needs, and determining the optimal moment to interrupt users without human intervention is an ongoing research area. For instance, Kern et al. identified five factors influencing user interruptibility: location, event importance, user's activity, social situation, and social activity~\cite{kern2003context}.

\subsection{Visual Cues}

Visual cues provide information to users through vision, ranging from simple error Light-Emitting Diode (LED) lights to complex AR notifications. Wallmyr et al. conducted a study on transparent interfaces using mixed reality to display key information to construction site operators. Their findings indicated that users responded more quickly and experienced lower workloads with head-up displays compared to head-down displays~\cite{wallmyr2019evaluating}.

The position of the visual cues is also important. Harrison et al. ~\cite{harrison2009locate} conducted a study to examine the reaction time performance based on the placement of visual cues, specifically small blinking lights on different body parts. Their study aimed to determine the effectiveness of visual cues in terms of accessibility, stability, comfort, social acceptability, and information conveyance. They found that reaction time performance varied significantly with the position of the lights, ranking from highest to lowest as follows: wrist, arm, brooch, shoulder, thigh, waist, and shoe. The performance was influenced by physical distance, visual accessibility, and external factors such as occlusion by furniture.

The study by Weber et al.~\cite{weber2016design} investigated the use of notifications on smart televisions (TVs), which are often shared by multiple users. Unlike other smart devices, the notification mechanisms for TVs cannot be personalized. Through three focus groups, the study gathered impressions regarding the duration, amount of information, position, and number of notifications. Based on these findings, the researchers highlighted that the notifications that are truly important should be displayed, privacy is important when multiple people use the TV, and the notifications should be displayed when there is a break.

\subsection{Audio Cues}

In situations where visual cues can be distracting and harmful, audio cues are recommended. Lee et al.~\cite{lee2004collision} studied a collision detection system with audio and haptic cues to mitigate driver distraction. They found that graded warnings (gradually increasing levels) were preferred and performed better than single-stage warnings. Graded warnings were also less irritating and more trustworthy. While user performance was the same for haptic and audio cues, users preferred haptic cues in terms of trust, overall benefit to driving, and annoyance.

Another way audio notifications can be used is as a substitute for visual cues when the user is visually disabled. Crommentuijn et al. conducted a study to test different ways of providing audio or haptic cues when an obstacle is in the way~\cite{crommentuijn2006designing}. All auditory displays improved object localization compared to silence. Continuous spatial sound and sequential discrete auditory cues proved the most effective. Whereas, echolocation and auditory looming were somewhat less efficient.

\subsection{Notifications in Mixed Reality}

Mixed reality devices, ranging from smartphones with a rear camera to immersive HMDs like Oculus Quest Pro~\cite{meta2023questpro}, are primarily output visual information. The key considerations for displaying notifications or interrupting users include where to display information, the acceptable level of occlusion in different contexts, the timing of information display, the behavior of the currently running application, and the interaction methods for the user.
 
On the other hand, notifications in Virtual Reality (VR) cannot use the same framework as other smart devices without breaking immersion. Zenner et al.~\cite{zenner2018immersive} proposed a framework where notifications are integrated into the VR scenario, such as a villager delivering a letter in a medieval setting or a drone in a futuristic one, based on the context and urgency of the notification. While this approach maintains immersion, it requires additional work for developers and standardization of notification priorities. Rzayev et al. conducted experiments on the effects of notification positions in VR and AR, finding that top-right placement increased workload and reduced comprehension, and motion negatively impacted comprehension~\cite{rzayev2018reading}. Imamov et al. conducted a study to determine the optimal placement of interfaces in 3D space~\cite{imamov2020display}. They discovered that central and central-low positions on the display were the fastest, whereas the top left position was the slowest. In terms of depth, the closest position (1m) was the slowest, while positions at 2m and 3m were the fastest for central and central-low placements.

Cidota et al. investigated how different types of notifications impact workspace awareness and task performance in an augmented reality (AR) setting~\cite{cidota2016comparing}. The study compared three conditions: no notifications, audio notifications, and visual notifications, where a remote user was instructing a local user by the use of these notifications. Key findings reveal that users prefer visual notifications over audio or no notifications. The authors noted that visual notifications caused less cognitive overload, possibly because the game's tasks already required visual attention. An audio signal would have forced participants to divide their attention between two cues (audio and visual) instead of focusing on a single modality (visual). Woodward et al. conducted an extensive systematic literature review of 140 peer-reviewed studies to assess the effectiveness of augmented reality (AR) in enhancing situational awareness (SA)~\cite{woodward2023analytic}. Key findings indicated a significant gap in the use of specific SA evaluation techniques in the majority of user studies, with only 19\% included such methods. While several studies analyzed the color and style of text in AR, there was a notable lack of research on the users’ SA, emphasizing the need for future studies to explore these areas to improve AR's efficacy in maintaining SA.

In a study by Lee et al., the efficacy of displaying AR notifications in various positions on an HMD during dual-task performance was investigated~\cite{lee2022investigating}. Results showed that notification location significantly impacted performance, with the top-left position causing the highest task load and slowest response time and middle positions yielding the lowest task load and fastest response time. The study concludes that middle positions are optimal for AR notifications in dual-task scenarios, as they offer visibility with minimal disruption. The importance of display location in AR notification design, especially for multitasking users, is emphasized. According to the authors, this position provides sufficient notification visibility while causing the least disruption to the main job. The study emphasizes how crucial it is to consider display location when creating augmented reality notifications, especially when users must carry out several tasks at once.

\begin{figure*}[!bp]
  \centering
  \includegraphics[width=0.8\linewidth]{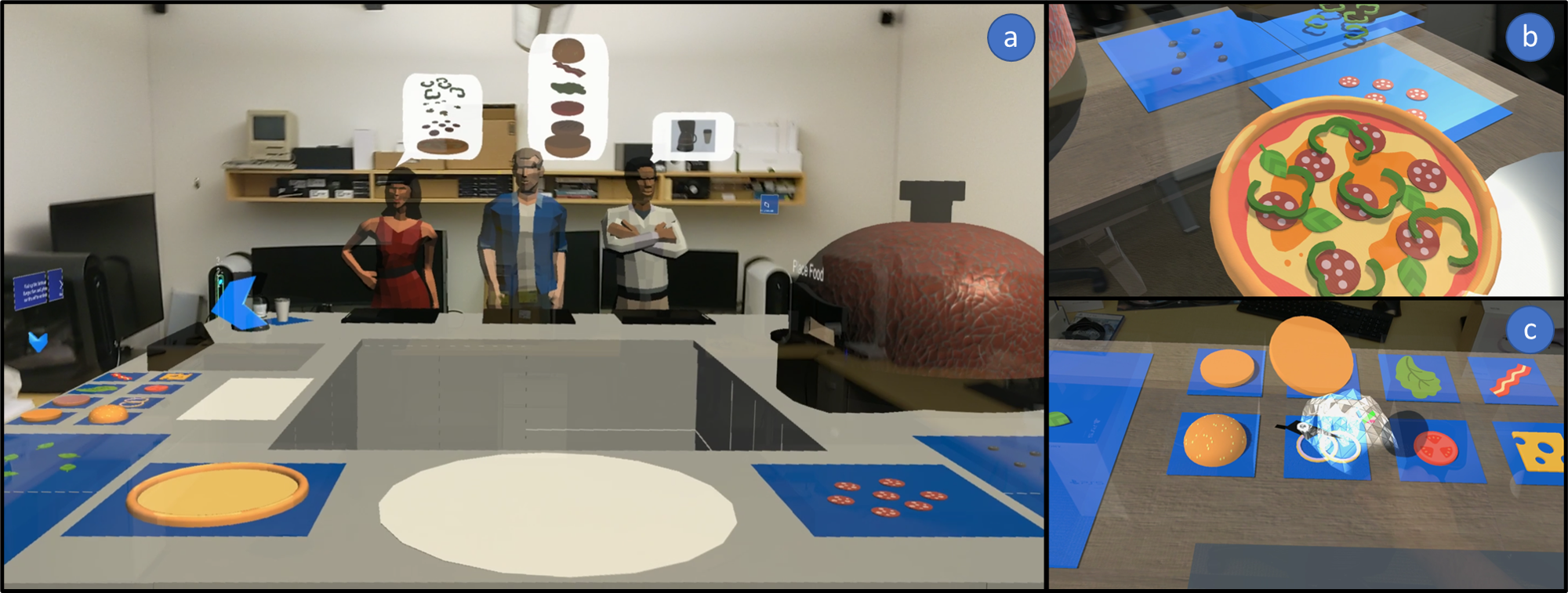}
  \caption{(a) Complete Setup of ARtisan Bistro; (b) Perfectly Cooked Pizza Taken out of the Oven; (c) Making Burger, Picking up Bottom Bun by Hand~\cite{raikwar2022artisan}}
  \label{fig:teaser}
\end{figure*}

Ghosh et al.'s NotifiVR extends notifications to visual, auditory, and haptic cues to alert users immersed in virtual environments to external events~\cite{ghosh2018notifivr}. They proposed five types of notifications for each cue type. For audio cues, they suggested that the sound played should be at the position of notification in 3D space,  metaphorical design like someone entering space as a door opening sound, and gradual increase of sounds to signify increasing urgency. For haptic cues, they recommended metaphorical vibrations like feet vibrating to signify footsteps, vibrations in the direction of notification, and a gradual increase of vibration to signify increasing urgency. For visual cues, they proposed notifications should be displayed in 3D space and not always attached to the viewport, metaphorical representations like clock face for alarms, blocks of notifications placed in 3D space, on the wall or surface, and in dangerous cases stopping the immersion and letting user reorient with the real-world space.

Plabst et al.~\cite{plabstPushRed2022} found that notifications in AR HMD are most effective when placed in the real world or at the bottom center of the headset's FoV, especially for sustained concentration tasks. Users also preferred these positions. Similarly, Lee et al.~\cite{leeExploringEffects2023} observed that bottom FoV placement resulted in higher noticeability and comprehension for both icon- and text-type notifications compared to top placement during an AR walking task. Lazaro et al.~\cite{lazaroInteractionModalities2021} recommend using both visual and auditory signals for AR notifications and suggest further research on notification placement.

The findings from Zenner et al.~\cite{zenner2018immersive} on the positioning of notifications informed our design choices, particularly in placing notifications in context-aware locations to enhance visibility and minimize disruption. Additionally, the study by Ghosh et al.~\cite{ghosh2018notifivr} on multimodal notifications and their contextual application provided a foundation for our exploration of integrating visual and audio cues effectively. Specifically, their recommendation to play audio cues at the position of the notification in 3D space was implemented in our study, further enhancing the relevance and impact of our notifications. By building on these foundational works, our results demonstrated that placing notifications directly on relevant objects can improve task performance and is preferred by users while also highlighting the crucial role of sound in capturing user attention.

\section {ARtisan Bistro}
\label{sec:artisanBistro}

ARtisan Bistro is an AR environment designed to simulate a fast food restaurant where users can prepare items like burgers, pizza, and coffee based on customer requests~\cite{raikwar2022artisan}. Developed using Unity engine v2022.1.10f1 and the Mixed Reality Toolkit (MRTK), the application serves as a tool for researchers to evaluate user interfaces\cite{artisan2023bistro}. The application was designed for researchers to evaluate user interfaces in general. We deliberately chose this open-source solution since it provides a standardized testbed for comparison of future results and replicates a familiar cooking scenario that is relatable to a wide audience.~\autoref{fig:teaser} shows different stills from the ARtisan Bistro environment.

\subsection{Environment Settings}
It is possible to change and set the environment in ARtisan Bistro. In this paper, we selected settings that simulate a high-stress environment, necessitating notifications to assist in task completion and promote multitasking across multiple food items concurrently.  

\textbf{Burger Station:} Participants were instructed to cook patties and assemble burgers with six layers: bottom bun, top bun, patty, and three random ingredients. The patty must be placed directly above the bottom bun, but the order of the other ingredients is flexible. The different cooking statuses for the patty were Uncooked, Cooked, and Burnt, with each status taking ten seconds to achieve.

ng of the level, the coffee pot's level resets to zero, and the coffee maker starts filling it automatically. The time to fill a cup was set to ten seconds of continuous pour.

\textbf{Customers:} Among the twelve meshes available for customers, three females and three males were selected at random. Except for the tutorial level, all levels consisted of 6 customers who waited for their requested food for only two minutes. The time remaining was displayed right in front of the respective customer.

\section{Notification Design}
\label{sec:notificationDesign}

The notifications were designed with two modalities: visual and audio. Depending on the level, participants received notifications that were visual, audio, both visual and audio, or none.

\subsection{Visual Notifications}

AR-HMD environment offers a dynamic way to display notifications using 3D space. Zenner et al.'s method integrates notifications into the scenario to maintain immersion, but it may be excessive as it limits notification types and relies on a person to send messages~\cite{zenner2018immersive}. This method also places the responsibility of setting the correct priority level on the sender, which can vary among individuals. To address these issues, two types of visual notifications were designed to handle more general types of notifications.

\subsubsection{Visual Notification Design and Properties}

\begin{figure}[htb]
  \centering
  \includegraphics[width=0.9\linewidth]{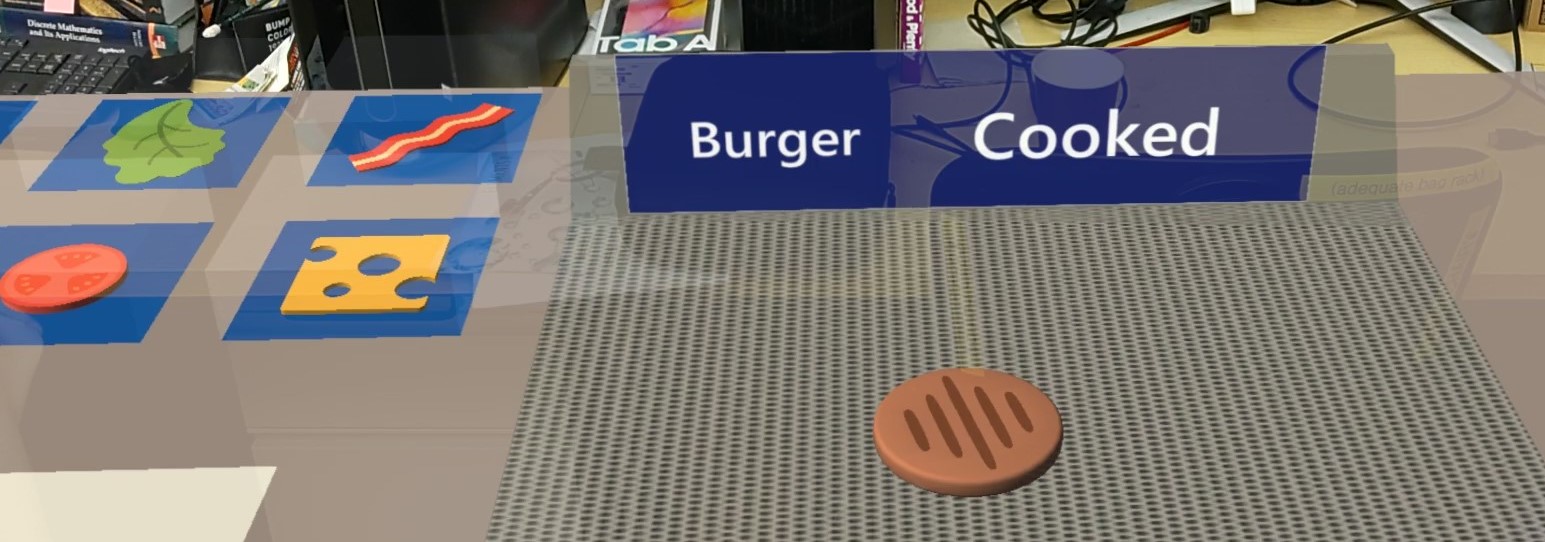}
  \caption{Notification Button}
  \label{fig:notificationBtn}
\end{figure}

The notification system, as illustrated in \autoref{fig:notificationBtn}, displayed information in a `Station Message' format. The `Station' indicated the cooking station's name, while the `Message' provided relevant information for that station. The notifications were cuboid-shaped, navy blue with white text. The notification design is based on the buttons provided and recommended by Hololens 2 developers. They were disabled after seven seconds. This duration, determined during the design phase of the experiment, was found suitable for users to pause their primary task, view the notification, and understand its content. Notifications that went unnoticed for seven seconds were disregarded.

The intervals between the notifications were based on the actions of users and the environment. For example, a customer stayed for 2 minutes after requesting their food. The next customer comes 5 seconds after the previous one leaves. In this case, the notification interval between these 2 customers will be 125 seconds, but if the participant serves the customer in 65 seconds, then the interval will be 70 seconds.

We used 4 texts to notify the participants. 

\textbf{New Customer:} When a new customer arrives requesting food.

\textbf{Cooked:} When either burger patty on the grill or pizza in the oven is cooked to the level of customer satisfaction.

\textbf{Burnt:} When either burger patty on the grill or pizza in the oven is cooked beyond the level of customer satisfaction.

\textbf{Coffee cup added:} When the coffee level in the coffee maker crosses a threshold where a complete coffee cup can be filled.

\subsubsection{Notification on Object}

\begin{figure}[htb]
  \centering
  \includegraphics[width=0.95\linewidth]{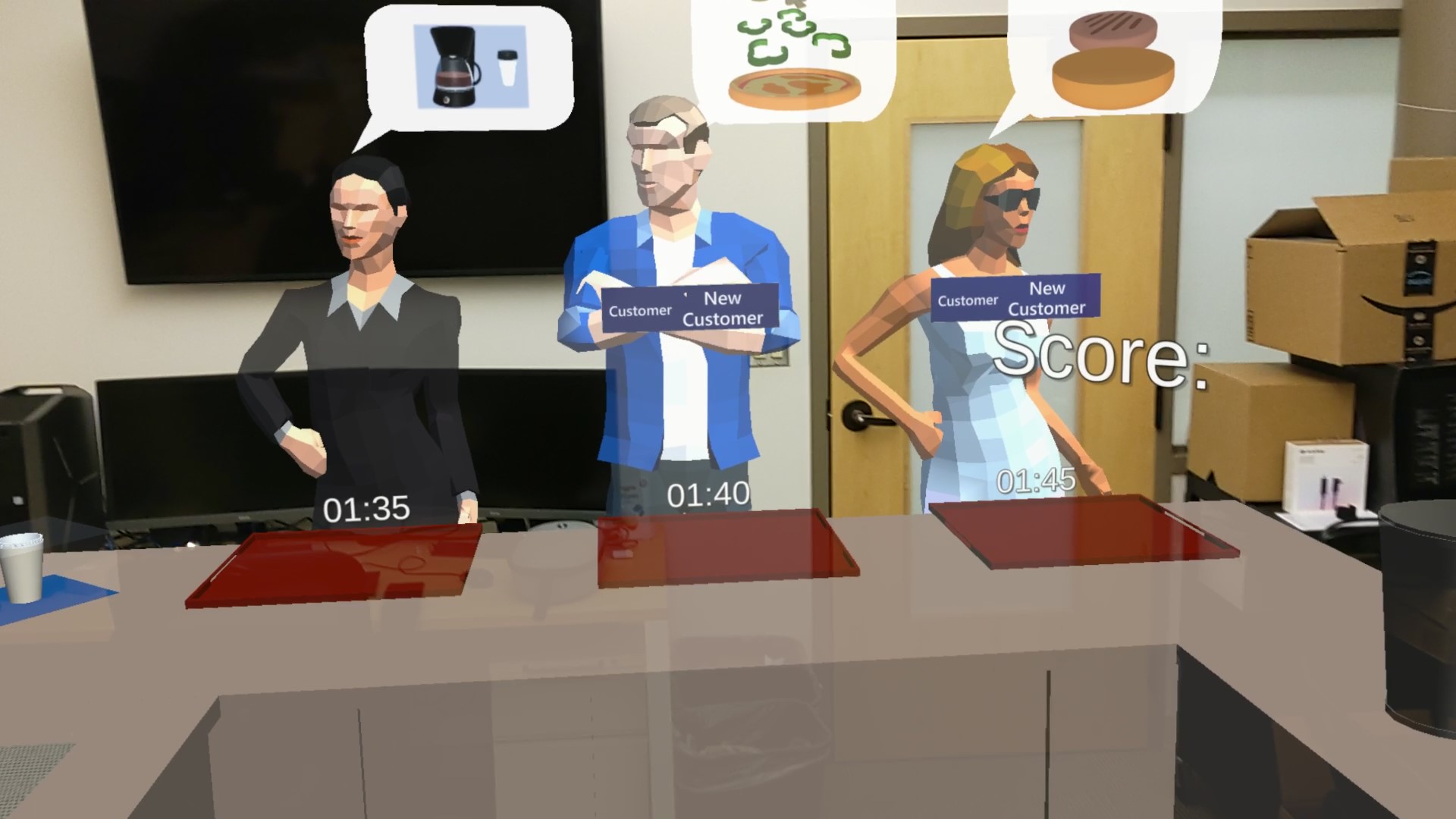}
  \caption{Notification on Object}
  \label{fig:notificationOnObject}
\end{figure}

Inspired by Zenner et al.'s~\cite{zenner2018immersive} implementation of immersive notifications, Notification on Object (\textit{O}) was created to present notifications in a context-aware 3D space. This approach aims to provide users with relevant information without overwhelming them, especially when focusing on specific tasks. For instance, in a cooking scenario, notifications about the pizza's cooking status can be displayed over the oven, allowing users to respond quickly without breaking immersion.

Although this solution reduces clutter and increases response time to notifications, it can also lead to missed notifications if they are outside the user's field of view or if the user does not look at them in time. One solution is to use audio alerts for new notifications. Another option is to list all notifications in one place, allowing users to periodically check the list for any new notifications.

\subsubsection{Notification on Dock}
\label{NoD}

\begin{figure}[htb]
  \centering
  \includegraphics[width=0.9\linewidth]{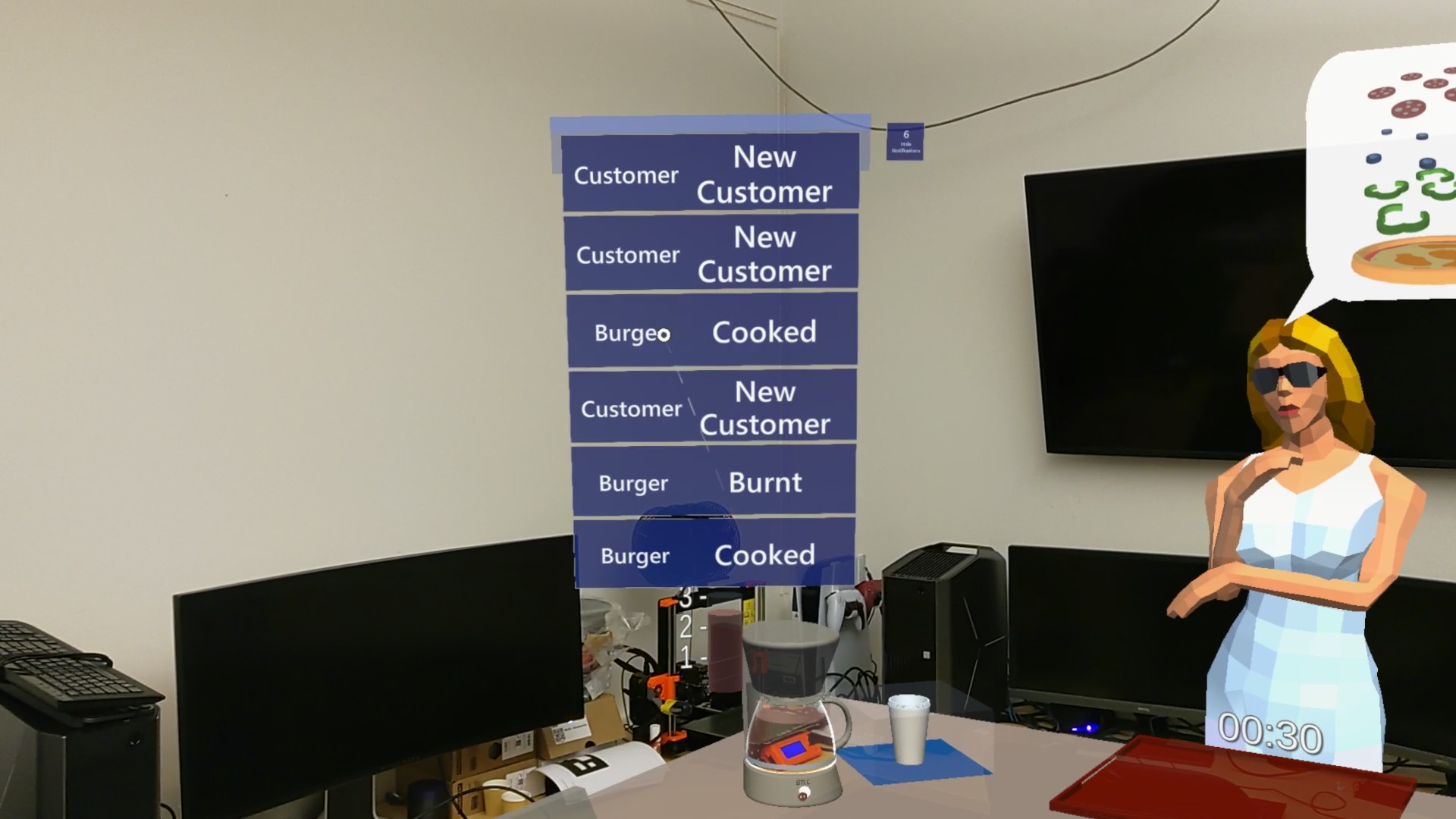}
  \caption{Notification on Dock}
  \label{fig:notificationOnDock}
\end{figure}

The concept of Notification on Dock (\textit{D}) is inspired by smartphone notifications, where all notifications are displayed in one accessible place by swiping down on the screen~\cite{google2024notification, apple2024notification}. This list format simplifies access to information and reduces the mental effort required to find notifications. Users might miss notifications if they appear randomly, but a centralized list allows users to periodically check for new notifications. Unlike object-based notifications, a dock cannot provide additional information beyond text (like the location of the object that prompted the notification), making it important to indicate the station that generated the notification.

Notification on Dock, like Notification on Object, has its drawbacks. Although it helps users locate notifications more easily, users must find an optimal position for the dock in 3D space. Inefficient placement can increase the time to find visual objects in 3D space~\cite{warden2022visual}. Additionally, the design does not inherently provide location information, necessitating extra details in the message to indicate the location.

\subsection{Audio Notification}
Incorporating audio cues serves as a potential solution to mitigate the limitation of visual cues not being feasible as the primary notification method~\cite{crommentuijn2006designing}. Different sounds can be assigned to different notifications, similar to how smartphones allow different tones for different applications. This helps users identify and prioritize notifications without needing to look at the application names.

In this experiment, only one localized sound clip played for notifications, allowing users to determine the direction of the sound. The Hololens 2's speakers, despite not offering a full surround sound experience, provide adequate left-right differentiation. Notifications were limited to the XZ plane, making the Hololens 2 speakers sufficient. Localized audio helped users identify the direction and source of the notification, such as different cooking stations, thereby reducing response time. The sound's origin depends on the notification's spawn location: on the dock for dock conditions and above the object for object conditions. In a condition with only sound and no visual notifications, the sound plays as a standard audio clip as if it were in the user's head, alerting participants to an event or task. There are two conditions for audio notification, with sound (S) and without sound (N).

\section{Methodology}
\label{sec:methodology}

\subsection{Research Questions}

Based on the previous work, we investigated the following Research Questions (RQ) in this paper: 

\textbf{RQ1} \textit{How does the inclusion of visual and audio cues influence the performance of users in serving customers?} Prior research shows that visual and audio cues can enhance user attention and reduce task load \cite{lee2022investigating, iqbal2010notifications}. By studying this in the context of serving customers, the research aims to understand how these cues help in managing interruptions and multitasking, potentially leading to more efficient and accurate task performance in AR, using MRT\cite{wickens2022applied}. Building on these studies, will users perform better in Notifications on Object with Sound~\textit{OS}, Notifications on Object without Sound~\textit{O}, Notifications on Object with Dock~\textit{DS}, Notifications on Object without Dock~\textit{D}, and no Visual Notification with Sound~\textit{S} conditions compared to the no Visual Notification without Sound~\textit{N} conditions?

\textbf{RQ2} \textit{What type of notifications are most noticeable?} The notifications should warn the user in a virtual environment when they focus on a task. With this RQ, we seek to determine which notification types are most prominent and immediately noticeable to users, ensuring that critical information is not missed. Which of these conditions (\textit{OS}, \textit{O}, \textit{DS}, \textit{D}) drew the user's attention the best?

\textbf{RQ3} \textit{What type of visual and audio cues do users prefer?} The previous work indicates that user preferences can significantly affect the perceived effectiveness and acceptance of notifications. For example, Lee et al. found that users preferred graded audio warnings in driving scenarios \cite{lee2004collision}. Which of the following conditions - \textit{OS}, \textit{O}, \textit{DS}, \textit{D} - was the most desired? By understanding user preferences in the context of AR-HMD notifications, designers can create more user-friendly and accepted notification systems that align with user expectations and needs.

These research questions are crucial for developing effective and user-friendly AR-HMD notification systems, ensuring that users can efficiently manage tasks and interruptions in dynamic and potentially stressful environments.

\subsection{Experiment Design}
\label{subsec:experiment_design}

This study used a within-subjects design with two factors: auditory notifications ($A_2$ =with audio notification and without audio notification) and visual notification ($V_3$ =notification on object, notification on dock, and no Visual notification), yielding 6 different conditions: Notification on Object Without Sound \textbf{\textit{O}}, Notification on Object With Sound \textbf{\textit{OS}}, Notification on Dock Without Sound \textbf{\textit{D}}, Notification on Dock with Sound \textbf{\textit{DS}}, no visual notification without sound \textbf{\textit{N}}, and no visual notification with sound \textbf{\textit{S}}.

Each participant followed the following steps. The experiment began with participants arriving at the research lab, completing a consent form, and receiving an explanation of the study. They watched a brief video to familiarize themselves with the environment and notification types. After addressing any questions, participants were given a Hololens 2 headset and entered the AR environment. They started with an introduction screen and then proceeded to a tutorial level where they practiced making food items for three customers without a time limit. The ingredient list was consistent for all participants.

After serving three customers, a level was chosen from a shuffled list of conditions. The order of levels was randomized for each participant using a chi-square model. Instructions were provided at the start of each level to inform participants about the type of notification they would encounter. This process was repeated until all six conditions were completed, followed by post-surveys.

Each level followed the script where six customers (maximum three at a time) made requests for a food item. The customers were virtual, asking for virtual food. The participants were tasked to make the item in a set amount of time. The appliances were also virtual, presenting extremely low to no risk to the participants. This scenario consisted of timed tasks where the participants had to worry about not only the customers' patience (tolerance time limit) but also the cooking status of different food items. The cooking status notifications were provided to aid the participants with the tasks and keep track of the cooking status. The notifications were presented using different modalities and at different positions. The pictorial representation of the experiment flowchart is available for reference in the Supplementary section.

During the study design process, it was determined that six customers (two requesting each food type) would be used in this study. The number of ingredients needed for a burger or pizza remained consistent across all trials and participants. The order of food items requested was decided by sampling without replacement. Each customer experienced identical wait times for all food items, ensuring that each participant received the exact same wait times for each food item.

In \textit{O} and \textit{D} conditions, the visual notification stayed active for 7 seconds. In \textit{D} conditions, if a new notification from the same station appeared, it replaced the existing one. Notifications from different stations were stacked. The notification sound was a bubble-popping sound (refer to the supplementary video for an example), chosen for its lightness and lack of food metaphor. An additional number, randomly selected between 0 and 15 and unique among active notifications, was added to the notification text to note observation. Participants acknowledged the notification by calling out this number. The measurements recorded included if a customer departed before 120 seconds, customer waiting time, total number of customers served, notification recognition frequency, and reaction time.

The environment was engineered to subject participants to stringent time constraints while incorporating slight physical demand elements. This can be confirmed using the overall Raw NASA-TLX scores~\cite{hart2006nasa}. The participants completed the NASA-TLX survey after they completed all the trial conditions. The average overall workload experienced by participants was 60.64 (SD = 10.84). This is above average global workload score reported in~\cite{grier2015how}. The average raw scores are listed in~\autoref{table:rawTLX}. 

\begin{table}[!htb]
\centering
\caption{Raw NASA-TLX scores}
\label{table:rawTLX}
\begin{tabular}{lcc}
\toprule
Dimensions & Mean & STD \\
\midrule
Mental Demand & 67.31 & 19.52 \\
Physical Demand & 51.92 & 23.86 \\
Temporal Demand & 75 & 19.86 \\
Performance & 65.38 & 20.8 \\
Effort & 71.56 & 15.86 \\
Frustration & 32.69 & 17.22 \\
\bottomrule
\end{tabular}
\end{table}

\subsection{Participants}
The experiment involved 26 participants, equally divided between males and females, aged 23 to 70 (M = 32.73, STD = 11.9). 
Participants consisted of students and staff from Colorado State University, as well as people who were not affiliated with the university. 84.6\% of participants were either students or faculty. 
69.23\% of the participants reported that they had used at least one AR or VR device. Participants received compensation in the form of \$20. Among the participants, 47.6\% reported that they have worked in a restaurant in some capacity (this includes as a cook or a waiter). This study was approved by the university's internal review board.

\section{Results}
To analyze the data gathered in the experiment, we performed a two-way ANOVA. If the assumptions of the ANOVA were not met, the data was transformed using Aligned Rank Transform~\cite{wobbrock2011aligned}. If significant effects were found, we conducted a post hoc pairwise analysis using TukeyHSD, or in the case of transformed data, the ART-C procedure~\cite{elkin2021aligned}.

\subsection{Performance}

\begin{figure}[!htb]
  \centering
  \includegraphics[width=0.9\linewidth]{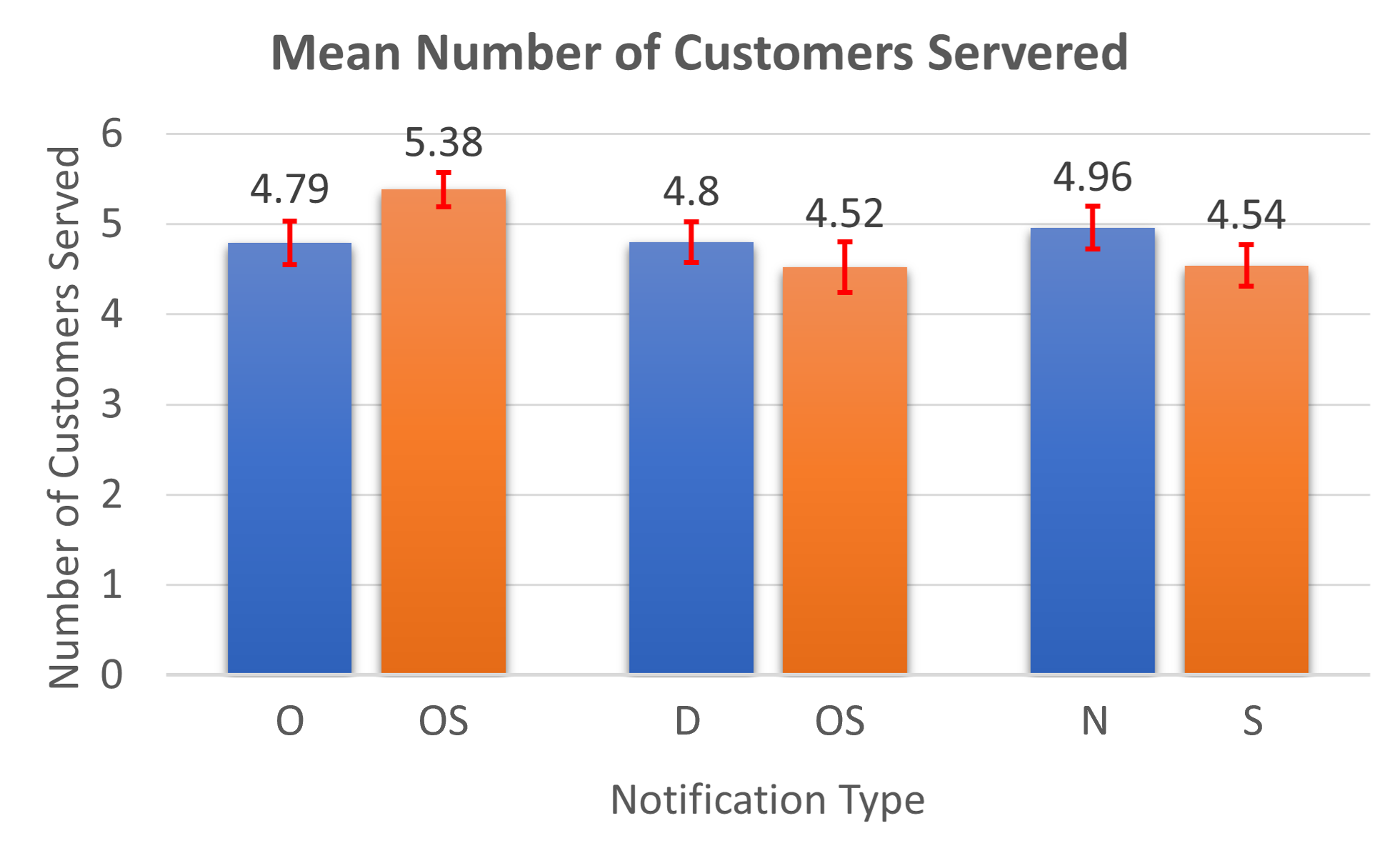}
  \caption{Performance of participants in different notification conditions based on how many customers they served.}
  \label{fig:mean_accuracy}
\end{figure}

Testing for the performance of the participants based on how many customers they served is shown in~\autoref{fig:mean_accuracy}. The x-axis represents the different conditions (3 conditions of visual notification and 2 conditions of audio notification). The y-axis represents how many customers were served on average in each condition (maximum is 6). The error bars represent standard error. Individual means, standard deviations, and standard errors can be found in~\autoref{table:meanPerformance}. During the ART-C procedure, the degrees of freedom are calculated using the Kenward-Roger Method~\cite{kenward1997small}.

We found that there was a statistically significant interaction between the effects of visual cues and audio cues ($F(2, 125) = 4.14$, $p < .05$). Simple main effects analysis showed that visual notifications ($F(2, 125) = 2.01$, $p = .14$) and audio notifications ($F(1, 125) = 0.27$, $p = .61$) did not have a statistically significant effect on the performance.

Contrast tests showed that Notifications on Object with Sound performed significantly better than both Notifications on Dock with Sound ($t(125) = 3.52$, $p < .01$) and Control with Sound ($t(125) = 3.16$, $p < .05$). There were no other significant differences found.

There was no significant effect on the performance of participants in serving the customers based on whether they have worked in a restaurant in some capacity ($t(24) = 1.6$, $p = .12$), despite participants with restaurant experience (M = 5.12, SD = 0.72) serving more customers than those without (M = 4.57, SD = 1.01). This was also true in the case of task load. NASA-TLX results show that there was no significant difference between the overall task load perceived by the participants who have worked in a restaurant (M = 50.15, SD = 13.33) compared to those who haven't (M = 49.62, SD = 11.56), $t(24) = 0.11$, $p = .91$. Just looking at the temporal demand ($t(24) = -1.07$, $p = 0.29$), the participants who have worked in a restaurant (M = 7.08, SD = 2.36) reported a similar level of temporal demand compared to participants who haven't worked in a restaurant (M = 7.92, SD = 1.61). 

\begin{table}[htb]
\centering
\caption{Mean performance of participants in different notification conditions}
\label{table:meanPerformance}
\begin{tabular}{lccc}
\toprule
Notification Techniques & Mean & STD & SE \\
\midrule
\textit{O} & 4.79 & 1.24 & 0.24 \\
\textit{OS} & 5.38 & 0.96 & 0.19 \\
\textit{D} & 4.8 & 1.14 & 0.22 \\
\textit{DS} & 4.52 & 1.44 & 0.28 \\
\textit{N} & 4.96 & 1.19 & 0.23 \\
\textit{S} & 4.54 & 1.18 & 0.23 \\
\bottomrule
\end{tabular}
\end{table}


\subsection{Visual Notifications Called}

\begin{figure}[htb]
  \centering
  \includegraphics[width=0.9\linewidth]{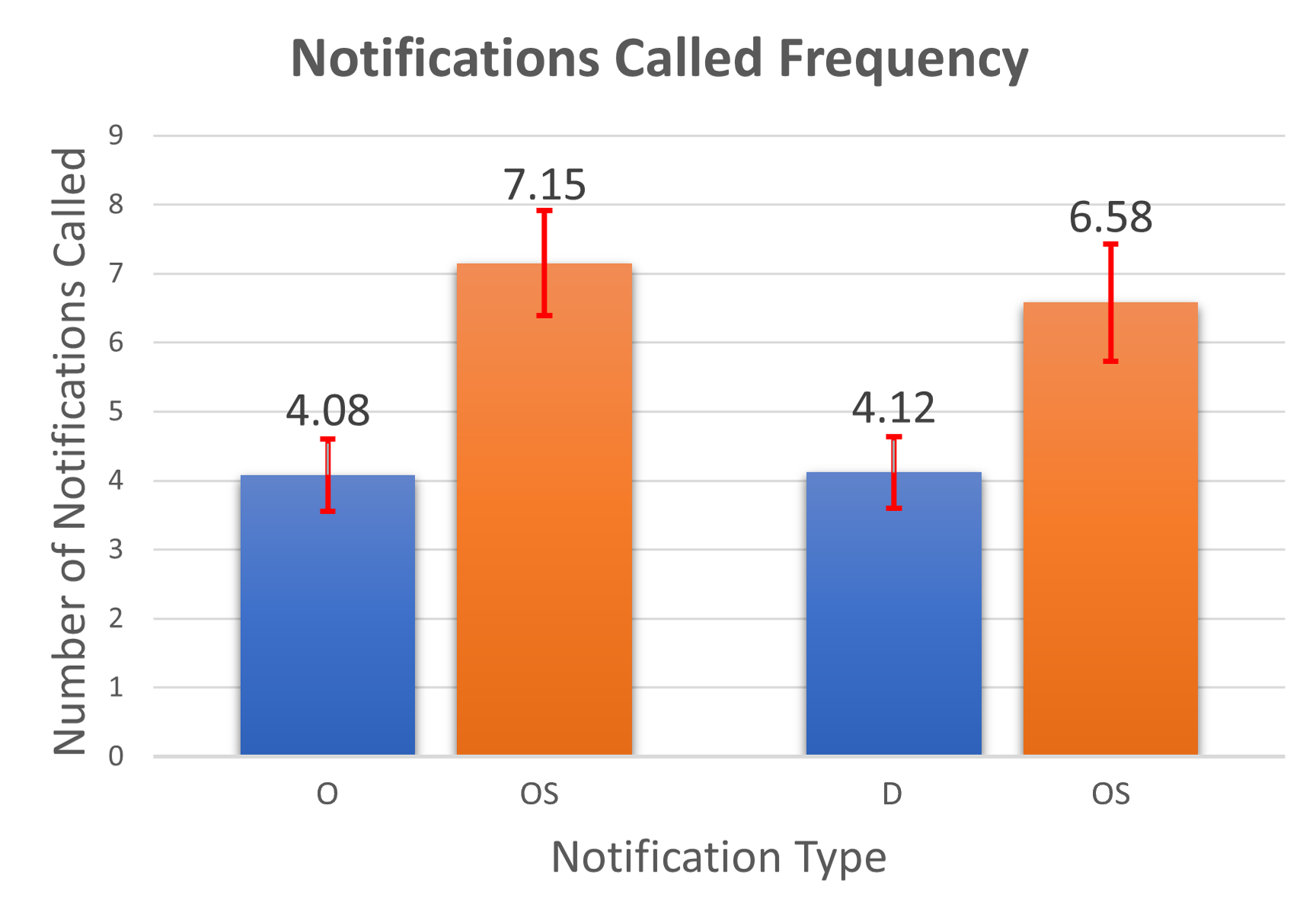}
  \caption{Average number of notifications called in different notification conditions}
  \label{fig:mean_notificationsCalled}
\end{figure}

As stated before, in addition to serving the customers, the participants were asked to shout out the number displayed above the text in a visual notification. Testing for how many visual notifications were noticed is shown in ~\autoref{fig:mean_notificationsCalled}. The x-axis represents the different conditions (2 conditions of visual notifications and two conditions of audio notifications). The y-axis represents how many customers were served on average in each condition (maximum is 6). The error bars represent standard error. Individual means, standard deviations, and standard errors can be found in~\autoref{table:mean_notificationsCalled}. 

Testing revealed that there was no statistically significant interaction between the effects of visual notifications and audio notifications ($F(1, 100) = 0.2$, $p = .66$). Simple main effects analysis showed that visual notifications did not have a statistically significant effect on noticeability ($p = .7$). Simple main effects analysis showed that audio notifications did have a statistically significant effect on the noticeability ($p < .001$), with more notifications being called out with sound (M = 6.865, SD =  4.18) than without (M = 4.1, SD = 2,71).

\begin{table}[htb]
\centering
\caption{Mean number of notifications called in different notification conditions}
\label{table:mean_notificationsCalled}
\begin{tabular}{lccc}
\toprule
Notification Techniques & Mean & STD & SE \\
\midrule
\textit{O} & 4.08 & 2.66 & 0.52 \\
\textit{OS} & 7.15 & 3.88 & 0.76 \\
\textit{D} & 4.12 & 2.65 & 0.52 \\
\textit{DS} & 6.58 & 4.32 & 0.85 \\
\bottomrule
\end{tabular}
\end{table}


\subsection{Visual Notifications Called Reaction Time}

\begin{figure}[!htb]
  \centering
  \includegraphics[width=0.9\linewidth]{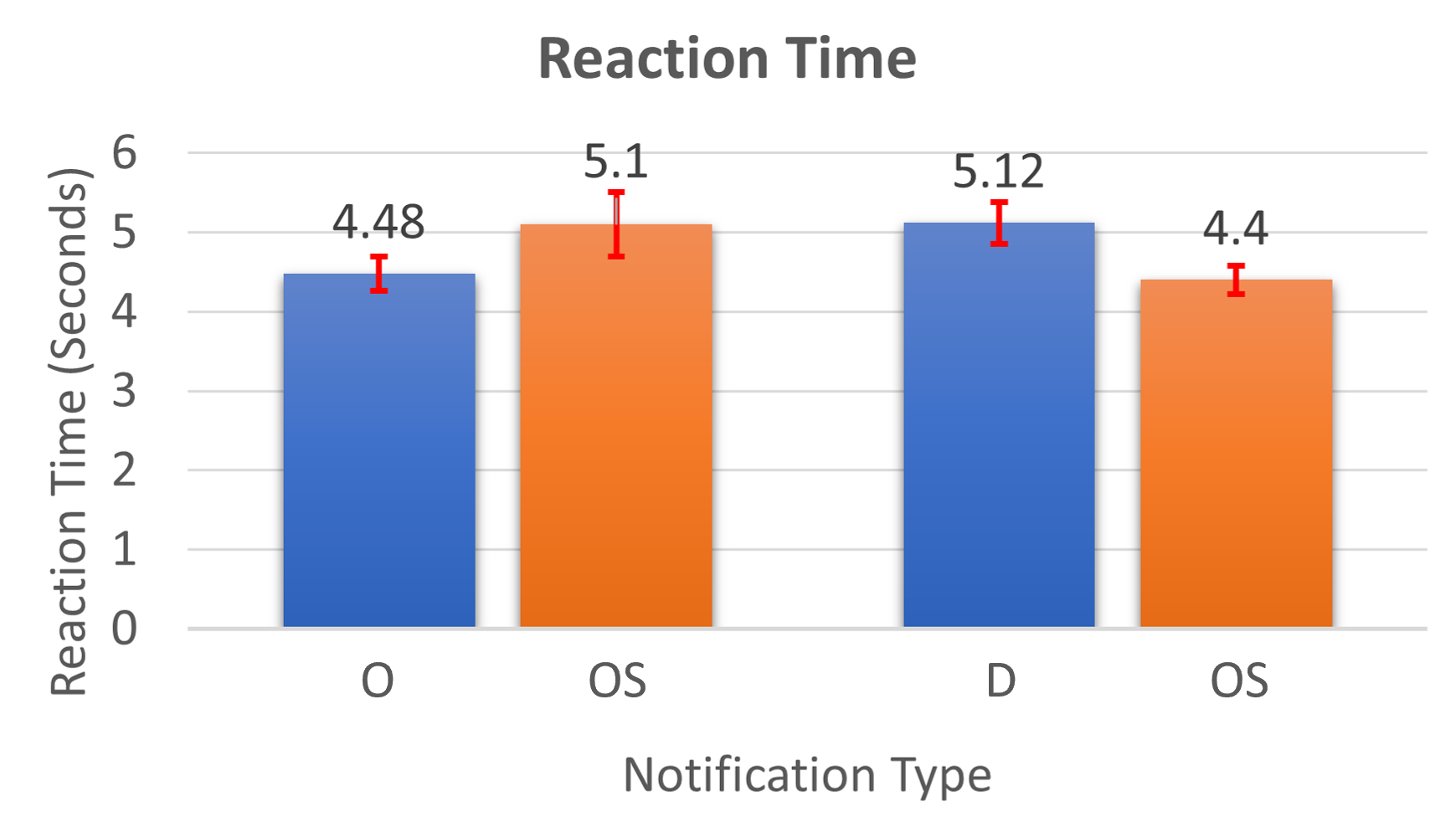}
  \caption{Average reaction time to notice the notification in different notification conditions in seconds}
  \label{fig:reactionTime}
\end{figure}

Testing for the reaction time of noticing visual notifications is shown in~\autoref{fig:reactionTime}. The x-axis represents the different conditions (2 conditions of visual notifications and two conditions of audio notifications). The y-axis represents the number of seconds the participants take to shout out the number on the notification. The error bars represent standard error. Individual means, standard deviations, and standard errors can be found in~\autoref{table:RT_notificationsCalled}. 

Testing revealed no statistically significant interaction between the effects of visual and audio notifications ($F(1, 68.08) = 3.28$, $p = .07$). Simple main effects analysis showed that visual notifications did not have a statistically significant effect on reaction time ($p = .36$), and audio notifications did not have a statistically significant effect on reaction time ($p = .34$).

\subsection{User Preference}
The system's usability was assessed using the Post-Study System Usability Scale (PSSUQ)~\cite{lewis1992psychometric}. All participants completed the PSSUQ after completing all the conditions in the AR environment. There were 18 items that were scored on a Likert scale from 1 (strongly agree) to 7 (strongly disagree). Other than the 18 items, we added a preference question and asked participants which notification condition they preferred among 6 conditions (Notification on Object Without Sound, Notification on Object With Sound, Notification on Dock Without Sound, Notification on Dock With Sound, No Notification without sound, and No Notification with sound)

Overall, participants rated the system favorably on the PSSUQ, with a mean score of 1.23 (SD = 0.44), indicating generally positive perceptions of usability. After examining individual items, the participants agreed strongly with most of the statements. Given the subjective nature of the questionnaire, we have considered any mean agreement rate below 1.75 (25 \%) as strongly agreeing. Out of all the items presented in the survey, only item 4, which discussed task completion time (M = 1.85, SD = 1.38), and item 8, which talked about error messages (M = 2.38, SD = 1.88), had varying levels of agreement. For the complete survey and detailed results, refer to~\autoref{sec:supplemental_materials}.

The last question in the survey asked about the preferred notification type. The preferences were as follows: 50\% preferred Notification on Object With Sound, 7.7\% preferred Notification on Object Without Sound, 19.2\% preferred Notification on Dock With Sound, 0\% preferred Notification on Dock Without Sound, 19.2\% preferred No Notification With Sound, and 1\% preferred No Notification Without Sound.

\begin{table}[!htb]
\centering
\caption{Average reaction times to notice the notification in different notification conditions in seconds}
\label{table:RT_notificationsCalled}
\begin{tabular}{lccc}
\toprule
Notification Techniques & Mean & STD & SE \\
\midrule
\textit{O} & 4.48 & 1.1 & 0.21 \\
\textit{OS} & 5.1 & 2.07 & 0.41 \\
\textit{D} & 5.12 & 1.34 & 0.26 \\
\textit{DS} & 4.4 & 0.93 & 0.18 \\
\bottomrule
\end{tabular}
\end{table}

\section{Discussion}
\label{sec:discussion}

In this paper, we looked at how two distinct visual and audio notification systems affect the user's performance and preference. The visual notifications were strategically placed in fixed positions or directly above relevant objects, while audio notifications were localized to align with the visual notifications. Our results demonstrated that visual notifications positioned above objects and accompanying audio outperformed alternative configurations. Placing AR notifications in the world anchored to the task context (in our case, on the object) was also found to produce the quickest reaction time, as well as the highest preference and lowest task load in a study by Plabst et al.~\cite{plabstPushRed2022}, even though they did not feature audio. Notably, this combination of visual and audio notifications exhibited greater noticeability and was the preferred mode of notification by the participants.

These findings emphasize the importance of carefully choosing where and how visual and audio notifications are used in AR. In our discussion, we explored what these results mean and why they matter for designing AR interfaces.

Looking at how many customers (out of six) were served (RQ1), neither visual nor audio cues had any significant effect on their own, but the combination of the two showed promising results. We contend that the audio notifications prove effective in capturing users' attention while simultaneously asserting that visual notifications above the objects accelerate users' ability to locate and engage with their primary tasks. The post hoc analysis illuminated a noteworthy trend: when audio notifications are present, visual cues above objects aid the users significantly more than visual notifications on a dock and no visual notification. Intriguingly, our findings did not uncover significant distinctions between different visual cue configurations when audio notifications were absent. This observation may suggest that, in the absence of sound notifications, participants relied less on visual notifications as a primary means of task assistance.

This observation aligns with the outcomes of our noticeability assessment (RQ2), which showed that the noticeability of the notifications significantly improves when audio notifications are present. In a 3D environment characterized by distributed information, the likelihood of missing notifications outside the user's immediate FoV is substantial. Audio notifications, where feasible, offer an effective means to augment notification noticeability, corroborating a finding by~\cite{lazaroInteractionModalities2021} that audio notifications outperform visual notifications, but the combination of both performs significantly better than either. 

Investigating user preferences (RQ3), the results remained consistent with the previous results, with 50\% of the participants stating that they prefer notifications on objects with sound. Regarding the PSSUQ survey, the participants generally found the system to be usable and satisfactory. The slightly lower scores on task completion time might stem from the time-sensitive nature of the tasks, intentionally designed to be expedited. The reason for the low average score on the item discussing error messages could be attributed to participants only encountering an error message when presenting food to the customer. This message consisted solely of a brief text indicating an incorrect food item. Providing more specific feedback earlier in the task may improve the score. 

While further research is needed in the field of AR HMD notifications, our study has yielded valuable insights. We found that strategically placing visual notifications enhances user performance, with ``Notifications on objects with sound'' garnering the highest noticeability. Regarding audio notifications, even in environments devoid of ambient noise, the use of sound alone was effective in improving noticeability and was preferred by the participants.

\subsection{Design Implications}

The study found that while neither visual nor audio cues alone significantly affected user performance, the combination of both led to promising results. Developers should thus consider integrating both modalities into their AR applications to enhance noticeability and user engagement.

The study also highlighted the importance of the effective placement of visual notifications, particularly above relevant objects, to improve user task performance. This suggests that the developers should consider the placement of visual cues to improve users' ability to find and interact with objects in an AR environment. The sound cues also proved effective in capturing users' attention, and developers should explore incorporating sound cues where feasible, especially in 3D environments with distributed information.

\subsection{Limitations}
While our results favored the combination of visual and audio cues, it's important to underscore the significant role played by sound cues in guiding users' attention toward visual notifications. The design of a visual cue with attention-directing capabilities may yield intriguing outcomes. 

Another limitation of our study relates to the absence of ambient audio noise in the AR environment, which could potentially diminish the effectiveness of sound cues. Despite the presence of visual clutter and noise in the AR setting, the absence of concurrent auditory noise represents a distinct environmental constraint worth acknowledging.
Lastly, users had to call out notifications verbally when they saw them, for the experimenter to log the notice. This could introduce potential noise to the reaction times.

\section{Conclusion and Future Work}
\label{sec:conclusion}
In this study, we focused on the implementation of visual and audio notifications in Augmented Reality (AR) Head Mounted Displays (HMDs), recognizing their pivotal role in crafting user-friendly interfaces. We offered insights into the efficacy of different notification modalities through an empirical evaluation encompassing user performance and preference. Within the immersive context of ARtisan Bistro, an open-source AR sandbox environment, participants engaged in tasks involving cooking and customer service. We systematically manipulated the placement of visual notifications, exploring fixed positions versus locations above objects, and coupled these with localized audio cues.

Our findings unveiled a compelling narrative: the combination of visual notifications above objects, complemented by localized auditory cues, emerged as the most effective condition, outshining alternative configurations in AR HMDs. This approach not only improved user performance but also caught users' attention the most.

This study highlights the importance of judiciously selecting and situating visual and audio notifications within the augmented reality environment. We believe that our findings can help engineers, developers, and others who create 3D interfaces for AR applications consider and design effective notification systems. As we explore mixed-reality tech, these findings guide user-focused innovation for smoother everyday use.

In future research, we aim to examine more effective visual cueing techniques that not only serve as visual notifications but also guide users' attention toward these notifications. Additionally, we intend to transition from a fully virtual kitchen to a setting where physical objects are tangible and physical while retaining the notifications in a virtual form. Another potential area for further investigation is the exploration of different visual notification techniques, such as color-coding and positioning notifications in the FoV, to help users prioritize notifications. For instance, color-coding could be used to indicate the urgency or type of notification, allowing users to quickly assess the importance of incoming alerts. Different shapes, sizes, or animations could also be employed to differentiate notifications based on context or priority. Such enhancements could reduce cognitive load and improve user efficiency in managing multiple notifications in an AR environment.

\section*{Supplemental Materials}
\label{sec:supplemental_materials}

All supplemental materials are available for reference. These include (1) an Experiment flowchart, (2) the image of the play area, (3) Excel files containing the complete PSSUQ survey and its data and analyses, and (4) a short demo video.

\acknowledgments{This work was supported by National Science Foundation Awards 2327569 and 2238313, and Office of Naval Research grants N00014-24-1-2214 and N00014-21-1-2580.}

\bibliographystyle{abbrv-doi}

\bibliography{template}

\end{document}